# A Contribution of the HAWC Observatory to the TeV era in the High Energy Gamma-Ray Astrophysics: The case of the TeV-Halos[1]

Ramiro Torres-Escobedo[2*], Hao Zhou[2*], Eduardo de la Fuente[3*] for the HAWC Collaboration[4]

**2** Tsung-Dao Lee Institute, Shanghai Jiao Tong University, Shanghai, China
**3** Departamento de Física, CUCEI, Universidad de Guadalajara, México
**4** The full list of co-authors are presented at the end of this paper
\* Correspondig authors: torresramiro350@gmail.com, hao_zhou@sjtu.edu.cn, and eduardo.delafuente@academicos.udg.mx (speaker)

April 12, 2023

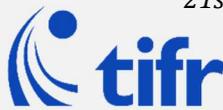

## Abstract

We present a short overview of the TeV-Halos objects as a discovery and a relevant contribution of the High Altitude Water Čerenkov (HAWC) observatory to TeV astrophysics. We discuss history, discovery, knowledge, and the next step through a new and more detailed analysis than the original study in 2017. TeV-Halos will contribute to resolving the problem of the local positron excess observed on the Earth. To clarify the latter, understanding the diffusion process is mandatory.

## Contents

---

[1]As part of the thesis to be submitted by Torres-Escobedo as a partial fulfillment for the requirements of Ph. D. Degree in Physics, Tsung-Dao Lee Institute, Shanghai Jiao Tong University. RT–E (Ph. D student). HZ (thesis advisor and director). EdelaF (external asesor).

# 1    Introduction

The history of gamma-ray astrophysics can be broken down into three important energy epochs: *high* (0.1-100 GeV), *Very High Energy* (0.1-100 TeV), and *Ultra-High Energy* (0.1 - 100 PeV). Concordantly, three main revolutions have happened on this topic in the 21st century: 1.- The GeV era ruled by the discoveries from FERMI-LAT (in the first decade of 21st century), 2.- The TeV era dominated by the maturity of the air imaging Čerenkov telescopes (IACTs), water Čerenkov detectors (WCDs), and plastic scintillator observatories mainly (∼ second decade of 21st century), and 3.- The PeV era unveiled by the rising of the PeV astronomy due to the high energy physics (HEP) like *PeVatrons* ( third decade of 21st century). Thanks to the development of the technology and the construction of suitable instruments and detectors in this century, it is feasible to perform studies impossible to do in the last century by combining observations from high-resolution telescopes and high-sensitivity observatories. For exam- ple, during its construction, the High Altitude Water Čerenkov (HAWC) observatory[2] (see §2) started to contribute to the TeV era with    one-third (from 106 to 133 WCDs; HAWC 111- stage; August 2013 to July 2014; livetime of 283 days) of its whole WCDs array (300 WCDs; HAWC-300 stage). The result was an unprecedented map of 2/3 parts of the gamma-ray TeV sky[3] with a median energy of 2 TeV, and a crab detection at a significance > $20\sigma$ [1, 2][4], overcoming in sensitivity to its predecessor: the MILAGRO observatory [3, 4][5]. This map was upgraded in 2016 by adding data from the HAWC-250 stage (live on November 26, 2014) to those in HAWC-111 stage and finally presented with data from November 2014 to June 2019 (a livetime of 1523 days) [5]. Another early result of HAWC was to present for the first time a small scale anisotropy map of cosmic rays[6,7] that finished into the first complete anisotropic cosmic-ray map combining HAWC (northern hemisphere) and Icecube (southern hemisphere); an example of a successful synergy of HAWC with other observatories [6,7]. Finally, as a com- plement, a scientific goal of HAWC is to study the all-particle energy spectrum of cosmic rays from 10 TeV to 1 PeV (see this proceedings [8] for details).

On the other hand, the extended TeV emission on Geminga was observed by MILAGRO with significance ∼$5\sigma$ in 2007 [9], and definitive detection in a ∼3° extension size with a significance of $6\sigma$ was confirmed two years later [4]. This result strengthened the idea of Geminga as a nearby cosmic accelerator [10, 11], able to explain the observed positron excess [12–14]. Nevertheless, the sensitivity of observations was not enough to perform a thorough study until combined with HAWC-111 and HAWC-250 data in 2017. Details and images on the Geminga analysis using HAWC data from November 26, 2014 to May 6, 2015 (live time of 149 days) with    $38\sigma$ are presented in [15, 16]. Thus, on the 2HWC catalog with    $500$ days of observations [17] with a 2° search, a significance of    $12\sigma$ and    $7\sigma$, HAWC detects two extended regions coinciding with Geminga and PSR B0656+14, respectively. As stated before, these pulsars, given their proximity to Earth, presented ideal candidates to investigate their contribution to the local positron excess (see §3.3). Nevertheless, these observations launched a new discussion of a new sub-class of gamma-ray sources and the reinterpretation of cosmic-ray diffusion within the galaxy.

This article provides an overview of the contributions of HAWC to the discovery of known TeV halos. §2 briefly introduces the HAWC Observatory. In §3, we briefly introduce the cur- rent interpretation of pulsar wind nebulae evolutionary stages, the characteristics of TeV halo candidates, and discussion about the relation between TeV-Halos objects and the local positron excess. In §4, we provide the concluding remarks and the expectations for future observations of new TeV halo candidates.

---

[2]https://www.hawc-observatory.org/

[3]See Figure 4 of [1]. At its latitude of 18.99° North, and considering 50° from zenith as the limit of the viewable field, HAWC observes up to 9 sr (> 70% of the entire sky) in a sidereal day.

[4]The HAWC main results from 2013 to 2015 were presented during the 34th ICRR meeting and proceedings

[5]During the lifetime of the experiment (2000–2008) MILAGRO detected the CRAB with a significance of $17\sigma$

[6]https://icecube.wisc.edu/wipac/2014/10/small-scale-cosmic-ray-anisotropy-with-hawc/

[7]https://wipac.wisc.edu/juan-carlos-diaz-velez-awarded-best-phd-thesis-from-centro-universitario-de-los-valles/

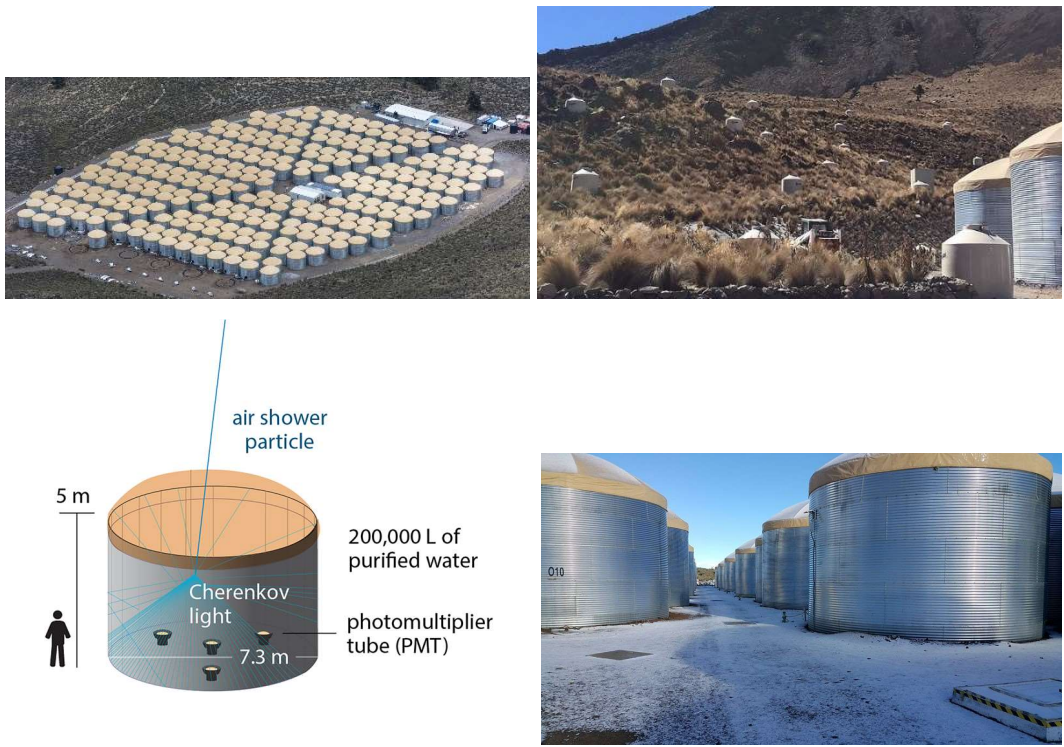

Figure 1: Top (left): The primary detector of the HAWC observatory consisting of 300 WCDs (array). Each WCDs houses 180, 000 litres of ultra-purified water. This array provides an effective detection area of 22,000 m². Top (right): A sparsely-spaced outrigger array of 300 smaller WCDs installed around the primary array. It is intended to compliment the primary array increasing its sensitivity by approximately four times. Bottom (left): HAWC primary detector WCD schematic. The arrangement of the PMTS is shown, with a ten-inches PMT surrounded by three eight-inches PMTs. Bottom (right): a true image of primary WCDs.

## 2 The HAWC Observatory

The HAWC Observatory was developed to observe the universe at energies between 100 GeV to 100 TeV. It is the successor of the MILAGRO Observatory. Thus, HAWC was built from experience gained from MILAGRO by reusing 900 eight-inch photo-multiplier tubes (PMTs) and electronics, plus extra components like the extra 300 ten-inch Hamamatsu PMTs (see §2.2). In context, it is essential to remark that a high sensitivity ground-based observatory does not conflict with a high resolution pointing telescope like IACTs but as a complement; justas in radioastronomy, a synergy between the sensitivity of a single dish, against the resolution by an interferometer. The aim is what kind of Science the scientist wants to do and what kind of Science the instrument can get.

## 2.1 History

The history of how HAWC landed in México is presented in *How HAWC landed in Mexico*[8]. The announcement was made at the 30th International Cosmic Ray Conference in Merida, México, in 2007 [18]. The location of Sierra Negra, Puebla, México at 4,100 m.a.s.l was chosen be- cause of the infrastructure offered by the Large Millimetre Telescope (LMT) "Alfonso Serrano", and by some Mexican colleagues in MILAGRO suggested México as a site with the support of American collaborators [19]. The latter includes the agreement between the Universidad Na- cional Autónoma de México (UNAM), and the Mexican national laboratory "Instituto Nacional de Astrofísica, Óptica and Electronica (INAOE) in Puebla", in assertive collaboration with the Consejo Nacional de Ciencia y Tecnología de México (CONACyT), high energy physics pioneer scientist of México[6], Los Alamos National Laboratory (LANL) representing the Department of Energy of USA (DOE), and the National Science Foundation (NSF) of the USA through the University of Maryland (UMD). Thus, HAWC started as a USA-México collaboration, including several Mexican and USA institutions in a truly binational collaboration context. In 2018, with the inclusion of Poland, Germany, Shangai, and Seoul institutions, the collaboration evolved into a USA-México-European_Union-Asia (China and Seoul Korea) collaboration[7] (see at the end of this paper).

## 2.2 The Instrument

The central detector of HAWC (see Fig. 1 top-left) comprises 300 WCDs (7.2m in diameter and 5m in height), covering an effective detection area of 22,000 m$^2$. Each WCD (see Fig. 1 bot- tom) braces a bladder with 180,000 litres of ultra-purified water with four PMTs anchored at the bottom to detect the Čerenkov radiation from secondary particles produced in extensive air showers (EAS) intiated by primary cosmic-rays and gamma-rays when they interact with Earth's atmosphere[9]. The PMT configuration in each WCD is one ten-inches Hamamatsu PMT at the centre of the WCD surrounded by three eight-inches Hamamatsu PMTs from MILAGRO as described above. Finally, in 2018, the installation of 300 outriggers (water Capacity of 1100 m$^3$) around the HAWC central detector was installed to improve the area and energy by a factor of four times (see top left panel of Fig. 1).

The HAWC observatory was built in stages. Firstly, it was developed with prototype engineered arrays: nano-hawc and VAMOS. These arrays linked the experience obtained in MILAGRO, and the knowledge needed to assemble HAWC (construction in 2.5 years). By September 2012, VAMOS was dismantled, and HAWC-30 (from 30 to 77 WCDs) still worked as a prototype array [20]. Then, from 77 WCDs, HAWC evolved to a 95-WCD array, and in August 2013, HAWC-111 began operations in July 2014. In this stage, HAWC was 3-5 more sensitive than MILAGRO, providing the first results like those mentioned above. Details on design, operation and reconstruction, and analysis are presented in [21]. In November 2014, HAWC started taking data with 250 WCDs (HAWC-250), and on March 20, 2015, HAWC was formally inaugurated in a solemn ceremony. Two years later, the more relevant scientific re- sults of HAWC include: the first Science paper in 2017 (Geminga and Monogem) and the first nature paper (jets of microquasar SS403 as TeV accelerators) [22]. Finally, in 2018, an out- rigger array based on MILAGRO was installed around the HAWC primary detector. The array consists of 350 tanks (diameter of 1.55 m, and height 1.65 m) with an 8-in PMT ankered to the bottom of the tank. Because cosmic-ray showers in the TeV energy range have a footprint comparable to the central array, such showers cannot be fully contained. With the addition of an outrigger array, the effective detection area for HAWC is increased 4-5 times, improving the angular resolution and energy reconstruction by an estimate of four times.

---

[8]https://www.symmetrymagazine.org/article/how-hawc-landed-in-mexico
[9]https://www.mpi-hd.mpg.de/hfm/CosmicRay/Showers.html

After its complete assembly, the HAWC observatory has access to an instantaneous field of view of 2 steradians, and with a > 95% duty cycle, it can survey 2/3 of the sky daily. Therefore, HAWC is an instrument capable of observing transient and continuous emissions from high-energy cosmic and gamma-rays. Furthermore, with its large field of view, HAWC can study galactic and extra-galactic point objects and extended galactic sources. Lastly, with the addition of the outrigger array, HAWC is expected to increase its sensitivity by an estimated four times. The 3HWC catalog, with three times more data ( 1527 days) than the previous catalog; 2HWC) [5], highlights the amount of physics that can be achieved with HAWC. With its high sensitivity, the HAWC observatory can study extra-galactic sources such as Mk 421, Mk 501 and extended objects, which is ideal for studying TeV-halos and indirect dark matter studies (see §3).

## 3 TeV Halos and the Local Positron Excess

The previous HAWC study of TeV halo candidates, Geminga and PSR B0656+14, was published utilizing a total of 500 days of live time observations binned by the fraction hit PMTs as a simple energy estimation technique. Therefore, a follow-up study will be presented in a later dedicated publication using two energy estimator techniques, "Ground Parameter" and "NeuralNetwork", developed for detailed energy analysis in [23]. The term "halo" here refers to the presence of over-density electrons/positrons in a region where the dynamics are not dominated by its parent Pulsar Wind Nebulae (PWN in singular; PWNe in plural). Therefore, the particle energy density must be less than that of the surrounding interstellar medium (ISM), $\epsilon_e \lesssim \epsilon_{ISM}$ [24]. The trajectories of electrons/positrons within this region get disrupted by magnetic fields. Then, after a series of inverse Compton interactions with low-energy photon fields such as microwave background, infrared, and stellar light (ISRF), leaving a halo of gamma-ray emission results in the formation of an isotropic gamma-ray halo. These halos can extend for tens of parsecs. Further, such halos have been observed in regions of slow diffusion.

### 3.1 PWNe Evolution

PWNe evolution can be briefly summarized in the following three stages (see Fig. 2 and Fig. 1 from [24]). The central pulsar "kick" velocity is assumed to be in the left of the direction and an ISM density gradient in the upwards direction. The first stage is considered for pulsar ages of $t \lesssim 10$ kyrs. During this stage, the forward shock (FS) of the parent supernova remnant (SNR) is in a state of deceleration with the surrounding interstellar medium (ISM). Beyond the contact discontinuity (CD), the deceleration produces a reverse shock (RS) which begins to shock the inner regions of the SNR. The pulsar is relatively close to its birthplace with a wind nebula delimited by the wind termination shock(WTS) with the inner region of its SNR. The second stage occurs for ages $10$ kyrs $\lesssim t \lesssim 100$ kyrs. By this stage, the RS now crushes and disrupts the structure of the PWN. Note that the interaction might happen faster for regions with higher density. Then, after several reverberations, electrons/positrons begin to escape into its surrounding SNR shocking material and the interstellar medium. At this stage, the pulsar has moved considerably from its birth location and escapes its parent SNR. The last stage occurs after $t \gtrsim 100$ kyrs., the parent SNR becomes subsonic and merges with the surrounding medium. The pulsar has escaped and formed a bow-shaped nebula. Electrons/positrons are now free to escape into the interstellar medium. As they escape, their trajectory gets randomized by the presence of magnetic, and after with low energy, interstellar radiation fields such as microwave background, infrared and stellar light (ISRF) photons generate an isotropic halo.

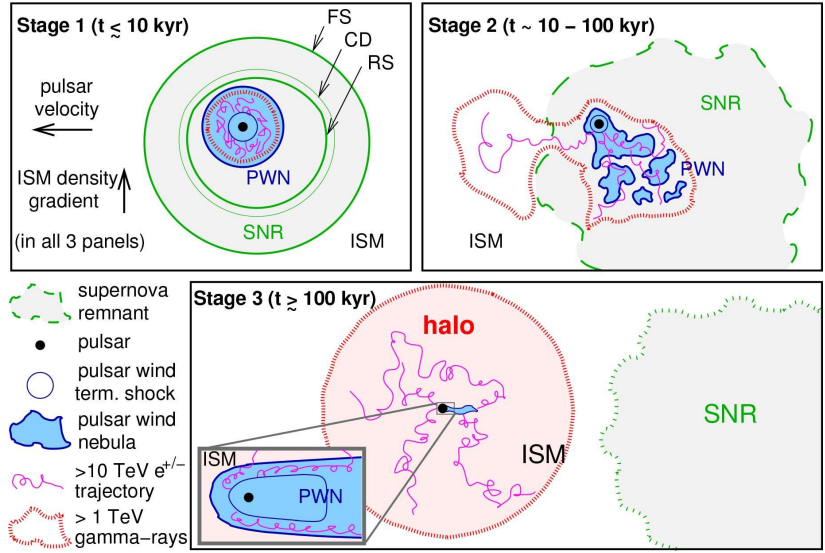

Figure 2: PWN evolutionary stages with emission shown for 1 TeV electrons/positrons. ISM density profile increases upward. The velocity of the pulsars is toward the left for all stages. During the first stage, TeV gamma-ray emission is constrained by WTS. The forward shock is in the deceleration stage with the surrounding ISM and generates a reverse shock beyond the contact discontinuity. In the second stage, the RS interacts first and disrupts the PWN, electrons/positrons escape into the shock SNR and the ISM. Pulsar escapes host SNR at the last stage, forming a bow-shaped PWN. Electrons/positrons generate gamma-ray halo from inverse Compton interactions (see Fig. 1 from [24]).

### 3.2 First TeV Halo Candidates

The first TeV halo candidates are Geminga, and PSR B0656+14 observed in the third stage. The HAWC collaboration studied the morphology of these sources with a one-zone diffusion model extending to Earth. Their morphology is consistent with a diffusion model (see Fig. 3) detected with $\sim 13\sigma$ and $8.1\sigma$, respectively [25]. The derived value of diffusion coefficient, $(4.5 \pm 1.2 \cdot 10^{27}$ cm$^2$/s), is significantly different from the average galactic value at 10 GeV ($D[10$ GeV$] \sim 8 \cdot 10^{28}$ cm$^2$/s). This discrepancy has led to the reinterpretation of particle diffusion within the galaxy. The LHAASO collaboration recently reported another TeV halo candidate associated with PSR J0622+3749 [26]. PSR J0622+3749 has a comparable age ($\sim$ 200 kyrs.) to that of Geminga and PSR B0656+14. The reported diffusion coefficient for this source is $D \approx 8.9^{+4.5}_{-3.9} \cdot 10^{27}$ cm$^2$/s, which is in agreement with the result of Geminga. The latter suggests slow-diffusion regions as a characteristic of TeV halos.

### 3.3 Local Positron Excess

An anomalous excess in the ratio of positrons to the total electron/positron flux, $e^-/(e^+ + e^-)$ first reported by PAMELA [12], and later confirmed by Fermi-LAT [27], and AMS-02 [28] to extend from few tens of GeV up to a few hundred GeV. This excess is incompatible with simulations of particle propagation. Although, the origin of this excess continues to be debated, a number of candidates have been postulated including PWNe [13,29], micro-quasars [30] and dark matter annihilation [31] to explain this local excess. Moreover, particle transport models in the ISM suggest that any source capable of accelerating very high energy electron/positron emission needs to be within a region of several kpcs [32]. The previous results from the HAWC study found no significant contribution to the local positron with a one-zone diffusion model and argued against this positron excess being the result of pulsars. Since then, a number of works suggest a two-zone diffusion model can explain this excess in the positron fraction

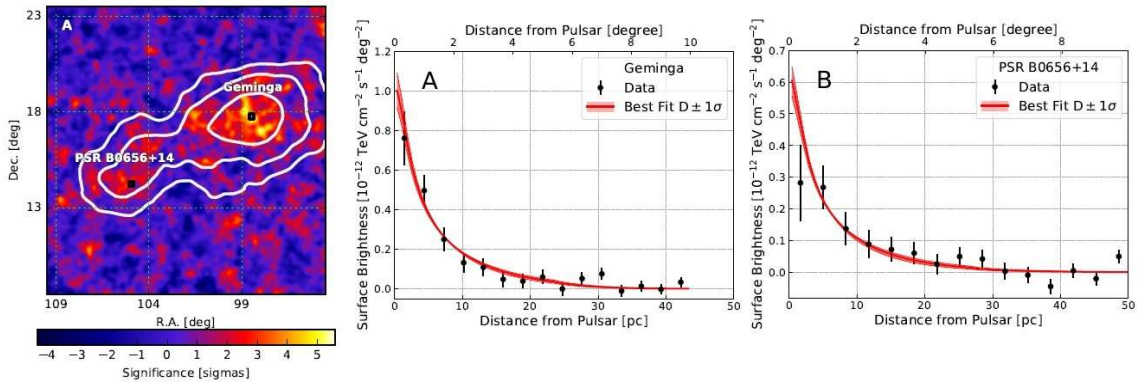

Figure 3: Left: Significance map from Geminga and PSR B0656+14 region. The white contours show the 5, 7, and 10$\sigma$ levels of detection. (see Fig. 1 from [25]). Surface brightness profiles from Geminga (center) and PSR B0656+14 (right). The red-line shows the best-fit value including the 1$\sigma$ uncertainty (Fig. 2 from [25]).

[33, 34], [35, 36] but more observations may be required.

## 4 Conclusion

With its wide field-of-view and high-duty cycle, the HAWC observatory provides an ideal instrument for the study of the continuous and transit study of point sources (e.g. Markarians) as well as that of extended emission significant for pulsar halos (TeV halos) and indirect dark matter searches. This article has presented an overview of the contributions from HAWC including observations of pulsar halos, and the most current understanding of PWNe evolutionary stages. HAWC'S first published results constrained a diffusion coefficient much smaller compared to that of the average galactic value with a $\sim$ 500 day data set. With a data set of $\sim$ 5 years now available, it is possible to conduct a much more detailed analysis of the energy-dependent slow diffusion region surrounding these halo candidates. The physical mechanism for the slow diffusion regions around these pulsar halos is the subject of continuing debate. Alternate interpretations such as two-diffusion region model, an alternate quasi-ballistic diffusion model. Wide-field cosmic-ray observatories such as LHAASO, SWGO are imperative in the detection of future TeV halo candidates.

## Acknowledgements

The following link https://www.hawc-observatory.org/collaboration/ contains the HAWC general acknowledgements. EdelaF and RT-E thanks cuerpo académico PRODEP-SEP UDG-CA-499 for support and management during the lifetime of HAWC project. They also thank Prof. Akchurin, Nural and the staff of Texas-Tech University (TTU) for academic and financial support. We thank the anonymous referee for their comments that improve the manuscript.

# Full Authors List of the HAWC Collaboration

A.U. Abeysekara[48], A. Albert[21], R. Alfaro[14], C. Alvarez[41], J.D. Álvarez[40], J.R. Angeles Camacho[14], J.C. Arteaga-Velázquez[40], K. P. Arunbabu[17], D. Avila Rojas[14], H.A. Ayala Solares[28], R. Babu[25], V. Baghmanyan[15], A.S. Barber[48], J. Becerra Gonzalez[11], E. Belmont-Moreno[14], S.Y. BenZvi[29], D. Berley[39], C. Brisbois[39], K.S. Caballero-Mora[41], T. Capistrán[12], A. Carramiñana[18], S. Casanova[15], O. Chaparro-Amaro[3], U. Cotti[40], J. Cotzomi[8], S. Coutiño de León[18], E. De la Fuente[46,54], C. de León[40], L. Diaz-Cruz[8], R. Diaz Hernandez[18], J.C. Díaz-Vélez[46], B.L. Dingus[21], M. Durocher[21], M.A. DuVernois[45], R.W. Ellsworth[39], K. Engel[39], C. Espinoza[14], K.L. Fan[39], K. Fang[45], M. Fernández Alonso[28], B. Fick[25], H. Fleischhack[51,11,52], J.L. Flores[46], N.I. Fraija[12], D. Garcia[14], J.A. García-González[20], G. García-Torales[46], F. Garfias[12], G. Giacinti[22], H. Goksu[22], M.M. González[12], J.A. Goodman[39], J.P. Harding[21], S. Hernandez[14], I. Herzog[25], J. Hinton[22], B. Hona[48], D. Huang[25], F. Hueyotl-Zahuantitla[41], C.M. Hui[23], B. Humensky[39], P. Hüntemeyer[25], A. Iriarte[12], A. Jardin-Blicq[22,49,50], H. Jhee[43], V. Joshi[7], D. Kieda[48], G J. Kunde[21], S. Kunwar[22], A. Lara[17], J. Lee[43], W.H. Lee[12], D. Lennarz[9], H. León Vargas[14], J. Linnemann[24], A.L. Longinotti[12], R. López-Coto[19], G. Luis-Raya[44], J. Lundeen[24], K. Malone[21], V. Marandon[22], O. Martinez[8], I. Martinez-Castellanos[39], H. Martínez-Huerta[38], J. Martínez-Castro[3], J.A.J. Matthews[42], J. McEnery[11], P. Miranda-Romagnoli[34], J.A. Morales-Soto[40], E. Moreno[8], M. Mostafá[28], A. Nayerhoda[15], L. Nellen[13], M. Newbold[48], M.U. Nisa[24], R. Noriega-Papaqui[34], L. Olivera-Nieto[22], N. Omodei[32], A. Peisker[24], Y. Pérez Araujo[12], E.G. Pérez-Pérez[44], C.D. Rho[43], C. Rivière[39], D. Rosa-Gonzalez[18], E. Ruiz-Velasco[22], J. Ryan[26], H. Salazar[8], F. Salesa Greus[15,53], A. Sandoval[14], M. Schneider[39], H. Schoorlemmer[22], J. Serna-Franco[14], G. Sinnis[21], A.J. Smith[39], R.W. Springer[48], P. Surajbali[22], I. Taboada[9], M. Tanner[28], K. Tollefson[24], I. Torres[18], R. Torres-Escobedo[30], R. Turner[25], F. Ureña-Mena[18], L. Villaseñor[8], X. Wang[25], I.J. Watson[43], T. Weisgarber[45], F. Werner[22], E. Willox[39], J. Wood[23], G.B. Yodh[35], A. Zepeda[4], H. Zhou[30]

[1]Barnard College, New York, NY, USA, [2]Department of Chemistry and Physics, California University of Pennsylvania, California, PA, USA, [3]Centro de Investigación en Computación, Instituto Politécnico Nacional, Ciudad de México, México, [4]Physics Department, Centro de Investigación y de Estudios Avanzados del IPN, Ciudad de México, México, [5]Colorado State University, Physics Dept., Fort Collins, CO, USA, [6]DCI-UDG, Leon, Gto, México, [7]Erlangen Centre for Astroparticle Physics, Friedrich Alexander Universität, Erlangen, BY, Germany, [8]Facultad de Ciencias Físico Matemáticas, Benemérita Universidad Autónoma de Puebla, Puebla, México, [9]School of Physics and Center for Relativistic Astrophysics, Georgia Institute of Technology, Atlanta, GA, USA, [10]School of Physics Astronomy and Computational Sciences, George Mason University, Fairfax, VA, USA, [11]NASA Goddard Space Flight Center, Greenbelt, MD, USA, [12]Instituto de Astronomía, Universidad Nacional Autónoma de México, Ciudad de México, México, [13]Instituto de Ciencias Nucleares, Universidad Nacional Autónoma de México, Ciudad de México, México, [14]Instituto de Física, Universidad Nacional Autónoma de México, Ciudad de México, México, [15]Institute of Nuclear Physics, Polish Academy of Sciences, Krakow, Poland, [16]Instituto de Física de São Carlos, Universidade de São Paulo, São Carlos, SP, Brasil, [17]Instituto de Geofísica, Universidad Nacional Autónoma de México, Ciudad de México, México, [18]Instituto Nacional de Astrofísica, Óptica y Electrónica, Tonantzintla, Puebla, México, [19]INFN Padova, Padova, Italy, [20]Tecnologico de Monterrey, Escuela de Ingeniería y Ciencias, Ave. Eugenio Garza Sada 2501, Monterrey, N.L., 64849, México, [21]Physics Division, Los Alamos National Laboratory, Los Alamos, NM, USA, [22]Max-Planck Institute for Nuclear Physics, Heidelberg, Germany, [23]NASA Marshall Space Flight Center, Astrophysics Office, Huntsville, AL, USA, [24]Department of Physics and Astronomy, Michigan State University, East Lansing, MI, USA, [25]Department of Physics, Michigan Technological University, Houghton, MI, USA, [26]Space Science Center, University of New Hampshire, Durham, NH, USA, [27]The Ohio State University at Lima, Lima, OH, USA, [28]Department of Physics, Pennsylvania State University, University Park, PA, USA, [29]Department of Physics and Astronomy, University of Rochester, Rochester, NY, USA, [30]Tsung-Dao Lee Institute and School of Physics and Astronomy, Shanghai Jiao Tong University, Shanghai, China, [31]Sungkyunkwan University, Gyeonggi, Rep. of Korea, [32]Stanford University, Stanford, CA, USA, [33]Department of Physics and Astronomy, University of Alabama, Tuscaloosa, AL, USA, [34]Universidad Autónoma del Estado de Hidalgo, Pachuca, Hgo., México, [35]Department of Physics and Astronomy, University of California, Irvine, Irvine, CA, USA, [36]Santa Cruz Institute for Particle Physics, University of California, Santa Cruz, Santa Cruz, CA, USA, [37]Universidad de Costa Rica, San José, Costa Rica, [38]Department of Physics and Mathematics, Universidad de Monterrey, San Pedro Garza García, N.L., México, [39]Department of Physics, University of Maryland, College Park, MD, USA, [40]Instituto de Física y Matemáticas, Universidad Michoacana de San Nicolás de Hidalgo, Morelia, Michoacán, México, [41]FCFM-MCTP, Universidad Autónoma de Chiapas, Tuxtla Gutiérrez, Chiapas, México, [42]Department of Physics and Astronomy, University of New Mexico, Albuquerque, NM, USA, [43]University of Seoul, Seoul, Rep. of Korea, [44]Universidad Politécnica de Pachuca, Pachuca, Hgo, México, [45]Department of Physics, University of Wisconsin-Madison, Madison, WI, USA, [46]CUCEI, CUCEA, CUValles, UDG-CA-499, Universidad de Guadalajara, Guadalajara, Jalisco, México, [47]Universität Würzburg, Institute for Theoretical Physics and Astrophysics, Würzburg, Germany, [48]Department of Physics and Astronomy, University of Utah, Salt Lake City, UT, USA, [49]Department of Physics, Faculty of Science, Chulalongkorn University, Pathumwan, Bangkok 10330, Thailand, [50]National Astronomical Research Institute of Thailand (Public Organization), DonKaeo, MaeRim, Chiang Mai 50180, Thailand, [51]Department of Physics, Catholic University of America, Washington, DC, USA, [52]Center for Research and Exploration in Space Science and Technology, NASA/GSFC, Greenbelt, MD, USA, [53]Instituto de Física Corpuscular, CSIC, Universitat de València, Paterna, Valencia, Spain [54]Institute for Cosmic-Ray Research (ICRR), University of Tokyo, Japan (Sabbatical 2021).